\documentclass[aip,amsmath,amssymb, reprint]{revtex4-2}
\usepackage{graphicx}
\usepackage{amsmath}
\usepackage{amsfonts}
\usepackage{amssymb}
\usepackage{float}
\usepackage[colorlinks]{hyperref}
\usepackage{multirow}
\usepackage{bbold}
\usepackage{soul}
\usepackage{makecell}
\usepackage{minibox}
\usepackage{soul}
\usepackage{fancyhdr}
\usepackage{mathtools}
\usepackage{amsthm}
\usepackage{overpic}
\usepackage[pdftex]{pict2e}
\usepackage[dvipsnames]{xcolor}
\usepackage{tabularx}
\usepackage[draft,inline, nomargin]{fixme}
\fxsetup{theme=color, mode=multiuser}
\FXRegisterAuthor{nb}{annb}{\color{blue}NB}	%
\FXRegisterAuthor{sug}{asug}{\color{red}Suggestion}	%
\FXRegisterAuthor{aps}{anaps}{\color{green}APS}	%

\begin{document}

\title{Deep-Ocean Application-Specific Neutrino Experiment}

\newcommand{\Tohoku}{Research Center for Neutrino Science, Tohoku University, Sendai, Japan}
\newcommand{\UH}{Unviersity of Hawai`i at Manoa, HI, USA}
\newcommand{\LLNL}{Lawrence Livermore National Laboratory, Livermore, CA, USA} 

\newcommand{\TohokuGeo}{Department of Earth Science, Tohoku University, Sendai, Japan}
\newcommand{\UMD}{Department of Geological, Environmental, and Planetary Sciences, University of Maryland, College Park, MD, USA}
\newcommand{\CAS}{Center for Geoneutrino Research, Institute of Oceanology, Chinese Academy of Sciences, Qingdao, China}
\newcommand{\Stanford}{Stanford University, Stanford, CA, USA}
\newcommand{\Carnegie}{University of Idaho, Moscow, ID, USA}
\newcommand{\UR}{University of Rochester, Rochester, NY, USA}
\newcommand{\KEK}{High Energy Accelerator Research Organization (KEK),
Tsukuba, Japan}

\affiliation{\Tohoku}
\affiliation{\Stanford}
\affiliation{\UH}
\affiliation{\LLNL}
\affiliation{\CAS}
\affiliation{\TohokuGeo}
\affiliation{\UMD}
\affiliation{\KEK}
\affiliation{\UR}
\affiliation{\Carnegie}

\author{Takumi Araki}
\affiliation{\Tohoku}

\author{Simran Chauhan}
\affiliation{\Tohoku}

\author{Lyla Choi}
\affiliation{\Stanford}

\author{Brian C. Crow}
\affiliation{\UH}

\author{Max A. A. Dornfest}
\affiliation{\UH}

\author{John~Graham}
\affiliation{\UH}

\author{Misaki Hosoya}
\affiliation{\Tohoku}

\author{John~G.~Learned}
\affiliation{\UH}

\author{Viacheslav A. Li}
\email{vali@llnl.gov}
\affiliation{\LLNL}

\author{William F. McDonough}
\affiliation{\CAS}
\affiliation{\TohokuGeo}
\affiliation{\UMD}

\author{Takeru~Ohno}
\affiliation{\Tohoku}

\author{Takanobu Ono}
\affiliation{\Tohoku}

\author{Taichi Sakai}
\affiliation{\KEK}

\author{Jackson Seligman}
\affiliation{\UH}

\author{Nathan Sibert}
\affiliation{\UH}

\author{Shang-Wen Stradleigh}
\affiliation{\UR}

\author{David Vartanyan}
\affiliation{\Carnegie}

\author{Hiroko Watanabe}
\email{hiroko@awa.tohoku.ac.jp}
\affiliation{\Tohoku}

\author{Zhihao Xu}
\affiliation{\Tohoku}

\author{Jeffrey G. Yepez}
\affiliation{\UH}

\date{\today}

\begin{abstract}
This report introduces the concept, prototype design, projected costs, and scientific goals of a mobile experiment for detecting geoneutrinos originating from uranium and thorium decay chains in the Earth’s mantle. This will constrain the planet’s radiogenic heat production and unearth its geochemical makeup. This design of a deep-ocean mobile neutrino experiment, which is not mirrored by any active or planned experiments, supports physics and geoscience's goal of multi-modal data on the Earth’s internal composition and structure. 
Based on geoscientific studies, this design is expected to achieve a 50–100-fold reduction in crustal background compared to similarly sized continental detectors, thereby enabling direct measurements of mantle geoneutrinos.
The multiple stereoscopic projections enabled by the detector’s unique mobility can map spatial variations in heat-producing elements within the mantle. Beyond discussing the design, we report on our collaboration's most recent hardware developments in the active prototyping of this detector. 
We briefly highlight the potential multiuse and interdisciplinary nature of this detector. 

\end{abstract}

\maketitle

\section{Science Program: Synergy between Earth Sciences and High-Energy Physics}

\subsection{Mantle geoneutrinos --- chemical composition of the Earth --- why Ocean-Bottom Neutrino Detector}

Uranium and thorium serve as long-lived fuel sources for Earth’s internal engine, supplying the energy that powers the geodynamo, which in turn shields life on our planet. Forged in ancient stars and later incorporated into the forming solar system, these elements were ultimately passed on to Earth. The radiogenic elements that heat our planet from within are, in a real sense, residual stardust. Directly establishing how much of this stellar legacy remains sequestered inside Earth is a clear and fundamental question that sits at the intersection of particle physics and planetary science.

How much heat is generated within Earth by radiogenic decay remains a central unresolved problem in Earth science. The bulk of this heat comes from the decay of the isotopes potassium-40, uranium-238, and thorium-232, which emit antineutrinos (geoneutrinos) through beta decays in their respective decay chains. Measuring Earth’s geoneutrino flux provides the only direct method for constraining the planet’s radiogenic heat production \cite{mcdonough2025earth}.

The central objective of geoneutrino studies is to quantify the absolute value and spatial variation of the surface flux of electron antineutrinos and electron neutrinos—collectively termed geoneutrinos—originating from $\beta$ and $EC$ decays within the Earth. Once measured, this flux can be converted into an estimate of the abundance and spatial distribution of uranium and thorium inside the planet. 

Uranium and thorium are refractory elements (they condensed from the solar nebula at high temperatures and in the same chondritic ratios, as evidenced by meteorites that acted as the raw materials for planets) and lithophile (they form bonds with oxygen and are preferentially incorporated into Earth’s silicate shell rather than its metallic core). As a result, they are enriched in the mantle and continental crust, with up to about 40\% of their total inventory residing in the continental crust.

Segregation of the Earth core was controlled by the chemical behavior of elements. Elements with chemical affinities for iron and nickel partitioned into the gravitationally dense, immiscible metallic alloy that sink to form the core. Based on its size and seismological properties, it is recognized that the core ended up with a bulk density that is about 10\% less than that of iron-requiring elements that bond with iron only under reducing conditions (e.g., H, C, O, Si, and S).

During core formation, the composition of the metallic alloys that segregated to build the core largely prevented these elements from being incorporated in substantial amounts \citep{fischer2025earth}. Consequently, constraining Earth’s total budget of refractory elements places tight limits on acceptable compositional models of the planet and determines the amount of radiogenic heat available to power geodynamic activity (e.g. plate tectonics, volcanism, mantle convection, and geodynamo).

Mantle convection underpins many processes that may be unique to Earth, including crustal recycling, plate tectonics, a geodynamo that has operated continuously for four-plus billion years, a mantle that has been cooling over the same time span, and the existence of long-lived mantle structures (i.e., LLVP, Large Low Velocity Province, 1000 km tall structures anchored at the core mantle boundary \cite{lay2006post, garnero2008structure}).

The thermal history of the Earth remains poorly constrained \citep{driscoll2023new,korenaga2026thermal}. At the most fundamental level, we can ask whether the planet cooled rapidly from its initially high temperatures or instead experienced a slower, more incremental decline. Over the last few billion years, has Earth cooled at a rate of the order of 50–70 K per billion years, or at an even lower rate? If the deep interior lost a large fraction of its primordial heat relatively quickly (as a result of impacts and a weakly insulating lithosphere), then its present-day cooling behavior would be closer to an asymptotic profile.

Determining how much radiogenic fuel is stored within Earth enables us to better constrain the cooling history of the mantle. However, due to present uncertainties, models of the thermal evolution of the Earth still yield a wide spectrum of values for the amount of primordial and radiogenic heat remaining inside the planet. Earth is made up of two fundamentally distinct regions with starkly different properties. Much like the difference between a cast iron skillet and a ceramic trivet in a kitchen, the metallic core is an extremely efficient conductor of heat, while the overlying mantle acts as a strong thermal insulator. Consequently, Earth’s cooling history is governed by heating from below (the core), internal heat production (radioactive decay) and the rate at which heat escapes from the surface.

The rate of change in mantle temperature ($dT_{M}/dt$) is determined by the balance between the production of radiogenic heat within the mantle, $h_M(t)$ and the loss of heat from the core and surface, $Q_{core}(t)$ and $Q_{surf}(t)$, normalized by the specific heat of the mantle, C$_M$.

\begin{equation} \label{thermalevol}
\frac{dT_{M}}{dt} = \frac{h_M(t) + Q_{core}(t) + Q_{surf}(t)}{C_M}
\end{equation}

\noindent
The most up-to-date estimates for these values are: a surface heat flux of 46 $\pm$ 3 TW, a core heat flux of 10 $\pm$ 5 TW, a radiogenic heat production of the mantle of 13 $\pm$ 6 TW, and a mantle specific heat of 1100 $\pm$ 100 J kg$^{-1}$ s$^{-1}$ \citep{jaupart2015temperatures,mcdonough2020radiogenic}. Based on inferred concentrations of K, Th, and U in the continental crust \citep{rudnick2014composition}, its corresponding radiogenic heat production is $7 \pm 1$ TW (denoted $h_{cont}(t)$).

Because the continental crust does not participate in mantle convection or in transporting heat from the deep interior of the Earth, the remaining 39 TW (after subtracting the continental share) are the focus of ongoing debate. In particular, there is uncertainty about the relative contributions of primordial heat and radiogenic heating to the share of the mantle of the Earth's internal energy budget. Since these uncertainties (23\%, 50\% and 46\%, respectively) are correlated, the covariance term in the numerator of equation \ref{thermalevol} leads to a combined uncertainty of $>$100\%. Future geoneutrino measurements are anticipated to significantly reduce uncertainties ($\leq$15\%) in both $h_{cont}(t)$ and $h_m(t)$, making the core heat flux the primary unknown.

Earth has two sources of fuel: primordial and radiogenic. Primordial energy comes from the kinetic energy generated during accretion and the gravitational energy accompanying core separation. Radiogenic energy comes from natural radioactive decay. Earth's production of radiogenic heat is mainly from $^{40}$K, $^{232}$Th, $^{238}$U and $^{235}$U (i.e., 99.5\% in total), with a lesser amount from $^{87}$Rb and $^{147}$Sm (i.e., 0.5\%). Based on $>$150,000 samples, \cite{wipperfurth:2018} showed that Kappa ($\kappa$ = $^{232}$Th/$^{238}$U) in mantle-derived basalts, crustal rocks, and chondrites was approximately 3.5 $\pm$ 1.0, and Kappa$_{Pb}$ ($\kappa_{Pb}$), which is derived by calculating the $^{208^*}$Pb/$^{206^*}$Pb value after subtraction of the Earth's initial $^{208}$Pb/$^{204}$Pb and $^{206}$Pb/$^{204}$Pb values and time integrating the contributions from the two decay chains. The values of the continental crust $\kappa^{CC}_{Pb}$ = 3.95$^{+0.19}_{-0.13}$ and the modern mantle $\kappa^{MM}_{Pb}$ = 3.87$^{+0.15}_{-0.07}$ bracket the initial solar system value of $\kappa^{SS}_{Pb}$ = 3.890 $\pm$ 0.15 \citep{blichert2010solar,wipperfurth:2018}. The value of $\kappa$ for the bulk silicate Earth is $\kappa^{BSE}_{Pb}$ = 3.87$^{+0.15}_{-0.07}$  (or Th/U =~3.77 for the mass ratio) \citep{wipperfurth:2018}. Currently, we cannot detect geoneutrinos in the K decay chain because their energies are below the threshold energy for detection \citep{wang2020hunting}. Geoscientists use the relative constancy of K/U 13,800 $\pm$ 1300 \citep{Arevalo2009} for mantle-derived basalts and crustal rocks to estimate the K content of 280 $\pm$ 120 of the bulk silicate Earth (BSE).

The total radiogenic power of Earth is estimated to be $ ( 19.9 \pm 3.0 ) $ TW, where 1 TW = $ 10^{12} $ W or J s$^{-1}$ of the heat generated within the Earth, according to a specific compositional model \citep{mcdonough2020radiogenic}. The predictions of competing compositional models span about 10 to 33 TW \citep{mcdonough2023neutrino}, indicating a factor of roughly three uncertainties in the absolute radiogenic power of the Earth. 
The present-day surface heat flux from Earth's interior is $ 46 \pm 3 $ TW \citep{jaupart2015temperatures}. The models, summarized in Table~\ref{tab:q_models}, vary widely in the relative contributions of primordial heat and radiogenic heating to this flux. There is general agreement that the continental crust contributes $ 7 \pm 1 $ TW of radiogenic power \citep{rudnick2014composition,mcdonough2023neutrino}. However, estimates of mantle radiogenic heat production remain highly uncertain, ranging from 2 to $ \geq 20 $ TW \citep{mcdonough2020radiogenic}. Thus, the abundances and distributions of heat-producing elements (HPEs), principally U, Th, and K, remain insufficiently constrained to tightly resolve Earth's thermal evolution.

\begin{equation}
\underbrace{Q_{\mathrm{total}}}_{46 \pm 3~\mathrm{TW}}
=
\underbrace{Q_{\mathrm{radiogenic}}^{\mathrm{crust}} }_{7 \pm 1~\mathrm{TW}}
+
\underbrace{Q_{\mathrm{radiogenic}}^{\mathrm{mantle}} }_{?}
+
\underbrace{Q_{\mathrm{primordial}}}_{?}
\label{eq:heat_budget}
\end{equation}

\begin{table*}[htbp]
\centering
\caption{Definitions of low-, middle-, and high-$Q$ BSE composition models, mantle, and crust combined. Here, ``Enstatite-chondrite-like'' refers to a model in which the Earth is assumed to have formed from meteorites with a particular chemically reduced composition, while ``chondritic'' refers more broadly to compositions similar to primitive, undifferentiated meteorites often used as a reference for the bulk composition of the Solar System.}
\begin{tabularx}{\textwidth}{|l|c|X|c|}
\hline
Model & Rad. heat estimate & Motivation & Ref. \\
\hline\hline
Low-$Q$ & 10--15 TW & Enstatite-chondrite-like mantle composition with low radiogenic element abundances. & \cite{javoy2010chemical} \\
\hline
Middle-$Q$ & 17--22 TW & Geochemical BSE estimate based on chondritic and terrestrial compositional constraints. & \cite{mcdonough1995composition} \\
\hline
High-$Q$ & $>$25 TW & Geodynamo requiring strong internal radiogenic heating to sustain mantle convection. & \cite{turcotte2002geodynamics} \\
\hline
\end{tabularx}
\label{tab:q_models}
\end{table*}

Electron antineutrinos, including geoneutrinos and those from nuclear reactors, are presently being detected by particle physicists in large detectors (0.3 to 20 kiloton), underground ($\sim$1 to 2 km deep). The total geoneutrino flux of the Earth is estimated to be approximately 5 $\times$ 10$^{25}$ $\bar\nu_e$ s$^{-1}$ or on average about 9 $\times$ 10$^{6}$ cm$^{-2}$~s$^{-1}$ \citep{mcdonough2020radiogenic}. Each neutrino leaving Earth removes a portion of its radiogenic heat.  The flux of geoneutrinos is proportional to the abundance and distribution of Th and U in the Earth. Despite their ghost-like nature, these particles, being nearly massless (i.e., $<$10$^{-6}$\;$\times \; m_e$ \citep{katrin2022direct}) and charge-less, are nearly impossible to detect, given their interaction cross-section of $<$10$^{-45}$\;cm$^2$ \cite{strumia2003precise}.
For $\mathcal{O}(0.1-1)$ MeV-scale reactor neutrinos and geoneutrinos, one would need approximately 1000 light years of lead to shield half of them.
This means that these particles, once generated, effectively leave the Earth with their complement of decay energy and traverse the universe.  Despite their elusive behavior, since 2005 particle physicists have detected about one electron antineutrino per 10$^{19}$ that pass through their detectors \citep{Araki:2005qa}.

\begin{figure}[h]
    \centering
    \includegraphics[width=\linewidth]{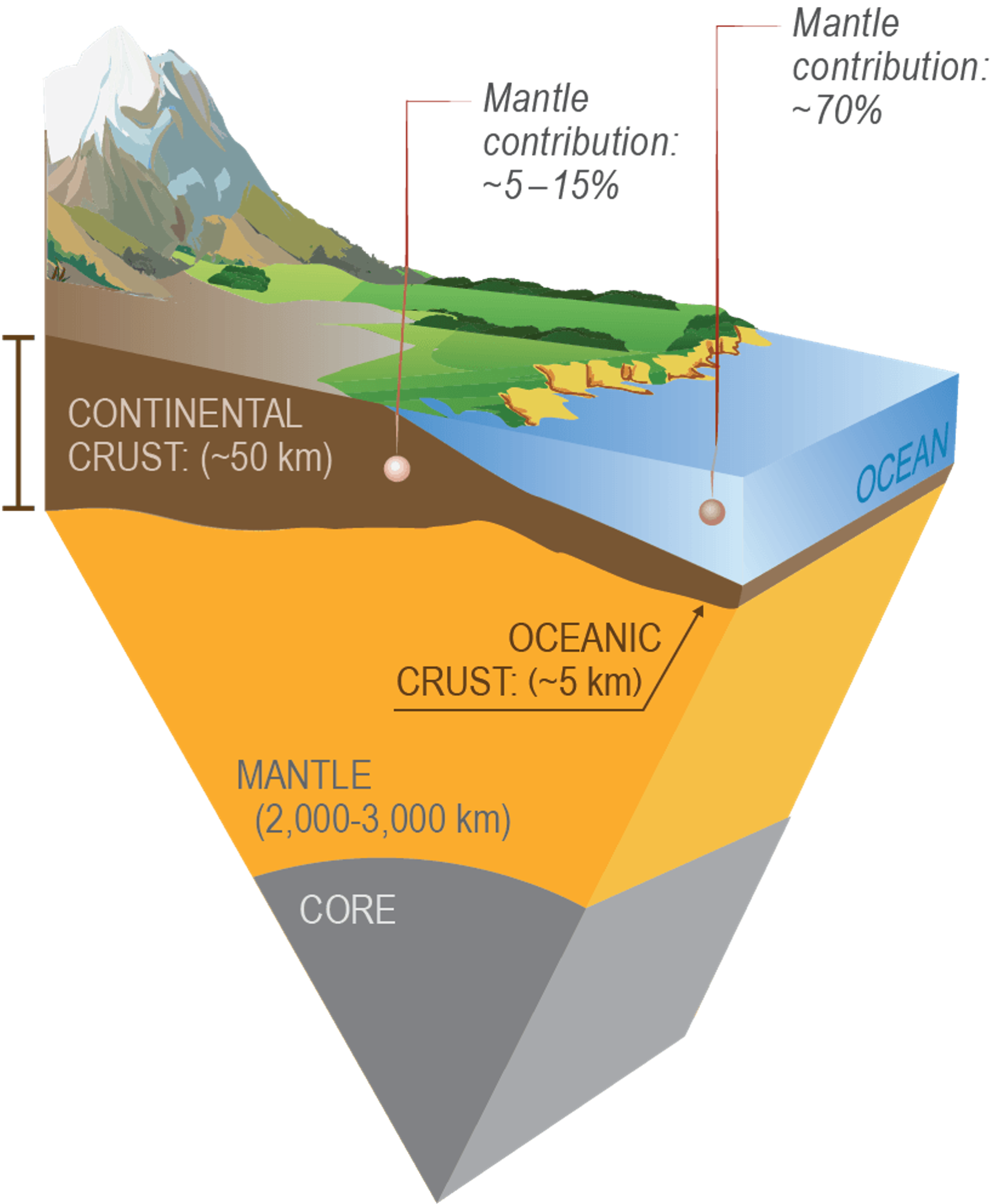}
    \caption{Artist’s representation (not to scale) of the signal contribution for a neutrino detector located on a continent versus a detector located in an oceanic region far from continents. The Earth’s cross section is not drawn to scale; in reality, the crust is much thinner than the mantle, comparable to the thin skin of an apple.}
    \label{fig_art}
\end{figure}

Today's detection technology for neutrino geophysics is strictly deployed on continental crust, with detectors sited either near the margins or deeper in the interior of the continents. Approximately 50\% of the signal measured at these detectors comes from the closest 250~km of continental crust \citep{mantovani2004antineutrinos,Araki:2005qa,coltorti2011u,huang2014regional,strati2015expected}, while the remainder typically comes in sub-equal proportions from the rest of the continents around the world and from the mantle. Consequently, translating the geoneutrino flux into a concentration of U and Th in the Earth requires an accurate and precise model for either the abundance and distribution of U and Th in the crust or the mantle. Knowing one of these parameters allows us to calculate the other. Therefore, our models for determining the concentrations of U and Th in the Earth depend entirely on the geological models that we build for both the mantle and the crust. If instead we were to measure a geoneutrino flux in an oceanic setting, far removed from the continents (e.g., southern equatorial Pacific, some 3000 km away from South America, Australia and the core-mantle boundary), we could assess the ``mantle only" flux from inside the Earth \cite{huang2013reference,wipperfurth2020reference}.

Following the first geoneutrino observation by KamLAND in 2005 \cite{Araki:2005qa}, subsequent geoneutrino measurement results were reported by Borexino in 2010 \cite{Borexino:2010dli}, SNO+ in 2025 \cite{SNO:2025koj}, and JUNO in 2026 \cite{abusleme2026measurement}.
The characteristics of these detectors are listed in Table~\ref{tab:geonu_detectors}.
All four detectors are kiloton-scale, liquid scintillator-based experiments located underground. In each case, the Earth's crust is the primary contributor to the observed geoneutrino flux. For instance, at KamLAND, approximately 70\% of the total geoneutrino flux originates from the crust, while the remaining 30\% is attributed to the mantle \citep{huang2013reference}.
All four detectors use the inverse beta decay (IBD) reaction, with neutron capture on hydrogen serving as the primary detection mechanism:
\begin{equation} \label{IBD}
    \bar\nu_e + p \to e^+ + n
\end{equation}

\begin{table*}
    \centering
    \begin{tabular}{|l|c|c|c|c|c|}\hline
        Experiment   & KamLAND & Borexino & SNO+ & JUNO & OBD \\\hline\hline
        IBD rate (TNU)          &         & $47.0^{+8.4}_{-7.7}\,(\mathrm{stat})^{+2.4}_{-1.9}\,(\mathrm{sys})$ & $73^{+47}_{-43}$ & N/A & \\\hline        
       Mantle total U+Th heat$^\ast$ (TW)         &  $10.6^{+5.2}_{-4.2}$  & $24.6^{+11.1}_{-10.4}$  &     N/A  &  N/A    & \\\hline  \hline
        Geoneutrino Signal (IBD events)       & 255 $\pm$ 49 & $52.6^{+9.4}_{-8.6}\,(\mathrm{stat})^{+2.7}_{-2.1}\,(\mathrm{sys})$ & $11^{+7.1}_{-6.6}$ & $83 \pm 24$ & \\\hline\hline
        Mantle signal / (mantle+crust)  (\%)       &  24
        & 20 &  19 & 21 & 74 \\\hline
        
        Mantle signal / total (\%)       &  15
        & 13 &  10 & 3 & 70\\\hline\hline
        Background total (events)   & 923 $\pm$ 35 & \(101.0\pm1.2\) %
        & 59 (obs) %
        & --- & \\\hline
        Background (reactor only)  & \(608^{+11}_{-13}\) & \(93.7\pm12.1\) %
        & $27.5 \pm 0.9$ (reactor) %
        & --- & \\\hline
        Observation time (years)    & 18  & 9 %
        & 1 & 1 & \\\hline
        First geoneutrino & 2002 & 2007 & 2023 %
        & 2025 &  \\\hline
        Exposure ($10^{32}$ proton-years)     &  $6.39 \pm 0.14$ & $1.29 \pm 0.05$  & $\sim 1$ (286 ton-years) & --- & \\\hline
        Target Mass (kt)        & 1 & 0.3 & 0.8 & 20 & 1--20 \\\hline
        Overburden (km.w.e.)   & 2.7  & 3.8  &  6 & 1.8 & 4--5 \\\hline
        Location     & Japan & Italy & Canada & China & Pacific Ocean \\\hline
        Reference         &  \cite{Araki:2005qa,Kamland_18years, TNU_KL_batygov_2006, TNU_KL_Borx_Fogli_2010} & \cite{Borexino:2010dli,PhysRevD.101.012009} & \cite{SNO:2025koj} &\cite{JUNOgeo2025,abusleme2026measurement} & \cite{Sakai2021DevelopmentOBD,Araki2024DevelopmentOBD,Ohno2024ComprehensiveSensitivityOBD, Chauhan2025OpticalPerformance,Ono2026DetectorStructureOBD,sakai2021study,watanabe2023ocean,xu2026proceedings}\\\hline
    \end{tabular}
    \caption{Summary of key parameters for liquid scintillator geoneutrino detectors (KamLAND, Borexino, SNO+, JUNO, and OBD). In these experiments, geoneutrinos are detected via inverse beta decay (IBD), with a threshold of  1.8~MeV. Where applicable, we list the measured geoneutrino rate in terrestrial neutrino units (TNU), signal and background yields, exposure, target mass, fiducial volume, overburden, and basic site information; quoted uncertainties are as reported in the cited references. Assuming homogeneous distributions of U and Th in the mantle.
    $^\ast$ For simplicity, in this table we only include ``Mantle total U+Th heat'' as measured with fixed U/Th ratio, this is mainly to highlight that the error bars are large regardless of the model used to separate mantle contribution; SNO+ has not acquired enough statistics yet to report a value.
    }
    \label{tab:geonu_detectors}
\end{table*}

KamLAND has collected data from March 2002 to December 2021 (a total of 5227 days, including 2590 days with reduced reactor background following the Fukushima-Daichi accident) \cite{Kamland_18years}. A comprehensive analysis of the geoneutrino signal from KamLAND is currently ongoing \cite{Kawada_NG2025}.
Borexino operated from 2007 to 2021 and published its most recent geoneutrino results in 2020~\cite{PhysRevD.101.012009}.
SNO+ is expected to update its geoneutrino measurement in the near future.
Another experiment, JUNO, a 20-kiloton liquid scintillator detector, started data taking in 2025 \cite{10.1088/1674-1137/ae3dc1}. In its first reactor antineutrino analysis, the geoneutrino contribution was included as a background component and extracted in the fit \cite{abusleme2026measurement}. However, a dedicated geoneutrino analysis, including detailed flux measurements, has not yet been reported. Owing to its large target mass, JUNO is expected to make significant contributions to future geoneutrino studies \cite{han2016potential}.

Shortly after the initial KamLAND observation, it was recognized that it is possible to probe the mantle directly via an ocean-deployed KamLAND-style detector. The concept was called Hanohano (Hawaii Anti-Neutrino Observatory), and was actively pursued by some of the members of our team~\cite{Learned:2007zz}.
Thus, this paper seeks, to a certain extent, to rejuvenate the original Hanohano idea and highlight the development over the past 20 years in this area. %

\begin{figure}[htbp]
    \centering
    \includegraphics[width=1.\linewidth]{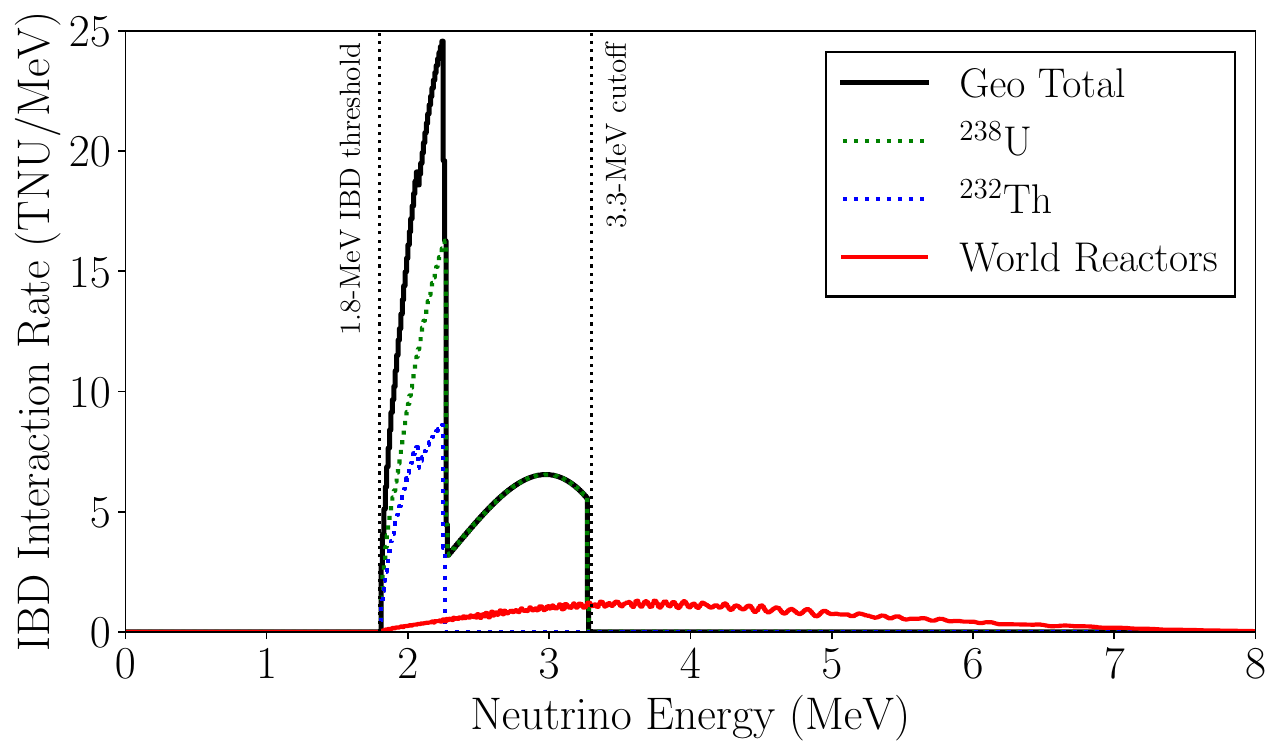}
\caption{Modeled antineutrino spectra for a mid-Pacific location, far away from continents and global reactors, in this case at the ALOHA deep-ocean site (22.8$^\circ$N, 158.0$^\circ$W). The spectra are shown as the expected inverse beta decay (IBD) interaction rate, obtained by folding the emitted $\bar{\nu}_e$ spectrum with the IBD cross section, and are plotted in units of TNU/MeV. The geoneutrino signal is dominated by mantle radioactivity (approximately 70\%), with the remainder from the crust. The contribution from commercial reactors worldwide is small at this location, particularly in the geoneutrino energy window, $ 1.8~\mathrm{MeV} < E_{\bar{\nu}_e} < 3.3~\mathrm{MeV} $; the high-energy endpoints are $\simeq 3.27~\mathrm{MeV}$ for the $^{238}$U chain and $\simeq 2.25~\mathrm{MeV}$ for the $^{232}$Th chain. Rates are given in terrestrial neutrino units (TNU), where 1~TNU = 1 event per $ 10^{32} $ free protons per year. 
}
    \label{fig_geonu_reactor_spectra_Hawaii}
\end{figure}

\subsection{Current state of affairs}

The original Hanohano concept included a high-energy physics program: deploying a kiloton-scale liquid scintillator detector at a 50--60\,km baseline offshore from a nuclear power complex to perform precision reactor antineutrino oscillation measurements. This medium-baseline program targeted improved determinations of $\theta_{12}$ and $\Delta m^2_{21}$ and sensitivity to the neutrino mass ordering using fine spectral structure. That HEP objective is now essentially being delivered by JUNO, a 20-kiloton liquid scintillator detector at a $\sim$53\,km baseline to the Taishan and Yangjiang reactors, designed for percent-level oscillation parameter precision and mass-ordering sensitivity. Consequently, the detector described here is application-specific: its primary mission is Earth science, namely isolating and measuring mantle geoneutrinos.

From a geoscience perspective, continental detectors have established geoneutrino observation and provided the first constraints on radiogenic heat production. 
However, at all continental sites the measured flux is dominated by the local and regional crust, and uncertainties in crustal composition and structure remain the leading systematic that limits inference on the mantle contribution \cite{sramek2016revealing}.

JUNO will accumulate a large geoneutrino sample due to its mass, but, like KamLAND and Borexino, it is located on continental crust and near powerful nuclear reactors, so crustal uncertainties and reactor backgrounds will continue to limit direct access to the mantle signal.
In contrast, a deep-ocean deployment at a mid-ocean site places the detector beneath oceanic crust that is both substantially thinner (typically 6--7 km versus 30--40 km for continental crust \cite{mooney2023earth}) and significantly depleted in heat-producing elements \cite{white2014}.
These factors reduce the crustal geoneutrino contribution by roughly a factor of 50--100.
Together with the large distance from nuclear reactors and the strong suppression of cosmogenic backgrounds at abyssal depths, this configuration is uniquely suited to isolating the mantle component of the geoneutrino flux.

The present concept refocuses Hanohano-like detector as a purpose-built, application-specific observatory for Earth’s interior: an ocean-bottom antineutrino detector optimized to measure the mantle geoneutrino flux and, by extension, Earth’s radiogenic heat budget.
Moreover, the mobility of a deep-ocean, relocatable detector enables measurements at multiple locations, creating a first-order map of heat-producing elements within the mantle \cite{Sramek2013}. 
In addition, advancing this technology provides the humankind with a mobile antineutrino detection capability that can be useful in other applications. Much like the benefits gained from sending humans to the Moon, we have reaped substantial rewards from being the first to develop such groundbreaking technology.

This experiment thus provides the first geochemical map of Earth’s deep interior and the only quantitative estimate of the current global amounts of Th and U inside the planet. 
Seismological data indicate that Earth’s interior is heterogeneous, featuring large-scale structures such as LLVPs, which are revealed by seismic tomography as regions of anomalously low seismic wave velocities beneath the Pacific Ocean and Africa relative to the surrounding mantle \cite{lay2006post, garnero2008structure}.
Their origin remains a subject of debate, with hypotheses ranging from the enrichment of heat-producing elements, such as uranium and thorium and accumulation of thermochemical material transported by downwelling mantle flow at the core-mantle boundary \cite{mcnamara2019review}, to Moon-forming impactor as a source of Earth’s basal mantle anomalies~\cite{Yuan2023MoonformingIA}.
By comparing geoneutrino fluxes originating from distinct mantle regions, the experiment can determine whether the spatial distribution of heat-producing elements aligns with seismic structures \cite{Sramek2013}.
Alternatively, such spatial variations can be resolved from a single observation site by exploiting the angular information of incoming geoneutrinos \cite{Xu2026hhx}.

\section{Prototyping and test sites in the US and Japan}
\subsection{Original Hanohano study}

Two decades ago, a detailed engineering study for the Hanohano deep‑ocean detector demonstrated that a 10,000 m$^3$ detachable, cylindrical anti‑neutrino instrument is technically feasible for mantle radiogenic heat measurements \cite{Learned:2007zz}. The preferred configuration is a portable detector cylinder mated to a separate support barge, with bathyscaphe‑like deployment driven by fluid buoyancy and a shore cable carrying fiber‑optic data and electrical power. Stability and staged filling were validated analytically and with a 1:50 pool model; ocean‑current envelopes and catenary cable‑lay were modeled; descent and ascent to roughly 4,000 m were each under an hour; and a practical Hawaiian shore landing was planned.

Key subsystems were prototyped and are directly applicable: oil‑buffered optical modules behind clear windows; localized high voltage at the sensor; neighborhood digitizers feeding fiber optics; and a simple tree network that sent all pulses to shore. Power came in at approximately 1~W per channel, with total system power under approximately 2~kW; battery‑only one‑year operation was marginal, so cabled power was preferred. Implosion risk was measured and mitigated, sympathetic implosions at close spacing were observed at sea, while inserting syntactic foam into instrument spheres reduced peak shock and prevented chain reactions. Communications at roughly 1 Gb/s over standard submarine fiber were well within practice. Costwise, the ship and barge portion accounted for about one fifth to one quarter of total project cost, a planning ratio that remains useful even though absolute amounts have changed.

\subsection{Tests at University of Hawai`i }
The ALOHA Cabled Observatory (ACO) is located approximately 100 km north of Oahu, Hawai'i \cite{howe2015aloha}. The observatory, which is actually a collection of instruments at Station ALOHA 4.7~km below the sea surface, established communications and power connection with Oahu providing oceanographic data on June 6th, 2011. The current maximum power rating is 1.2~kW. The observatory has hydrophones for listening to the ocean and a camera for visual observation. Various oceanographic measurements are constantly recorded, including the dissolved oxygen content, salinity, temperature and current profile of the water. The data is communicated to land via a retired AT\&T telephone cable between Station ALOHA and Oahu. The constant and long-term data from ACO will be analyzed to discover patterns of ocean circulation, ocean-atmosphere interactions, seismology, climate change and other oceanographic topics.

UH Manoa Physics is interested in a number of tests that are realizable through the ACO and on campus with oceanography engineering staff. In particular, they are able to run pressurized tests mimicking (and surpassing) the conditions at Station ALOHA. While seeking the funds to perform pressurized tests are ongoing, tabletop tests at ground level are being done on SiPMs (silicon photomultipliers) to test if SiPMs are a valuable new technology to incorporate in an ocean bottom detector. We are using a laser to photo-stimulate the SiPM and are interested in seeing what are compatible safeguards to increase the detector's survivability.

\subsection{Studies at Tohoku University}

Five M.S. theses at Tohoku University~\cite{Sakai2021DevelopmentOBD,Araki2024DevelopmentOBD,Ohno2024ComprehensiveSensitivityOBD, Chauhan2025OpticalPerformance,Ono2026DetectorStructureOBD} and three conference proceedings \cite{sakai2021study,watanabe2023ocean,xu2026proceedings} have advanced the feasibility of an ocean-bottom detector (OBD), addressing backgrounds, scintillator behavior at deep-ocean temperature, prototype optical performance, and deep-sea structural design. Below we summarize the main findings.

\subsubsection{Background evaluation} %

\begin{figure}[htbp]
    \centering
    \begin{overpic}[width=1.\linewidth]{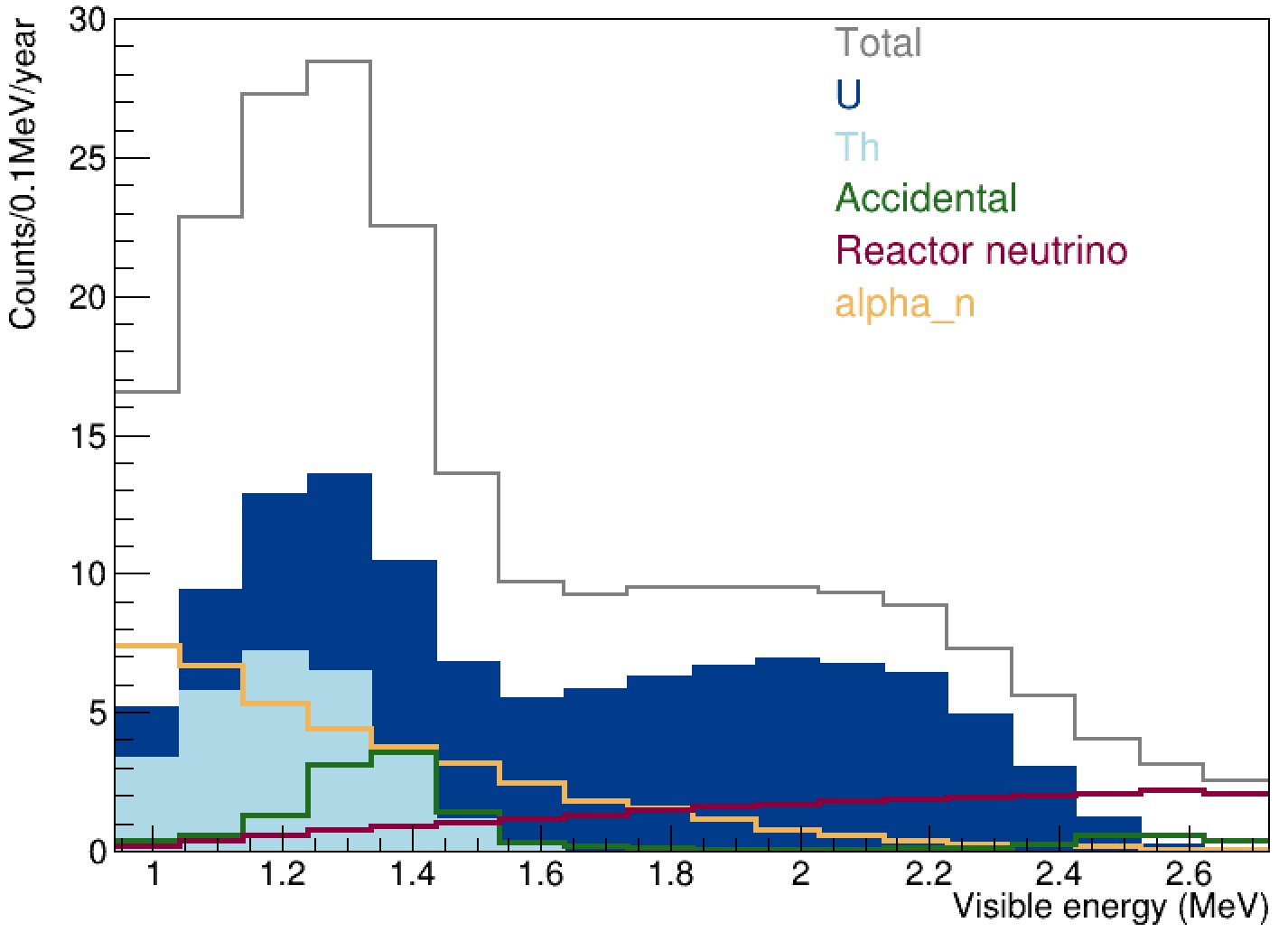}
        \put(50,37.7){%
            \setlength{\fboxsep}{1pt}%
            \setlength{\fboxrule}{0.6pt}%
            \fbox{%
                \includegraphics[width=0.45\linewidth]{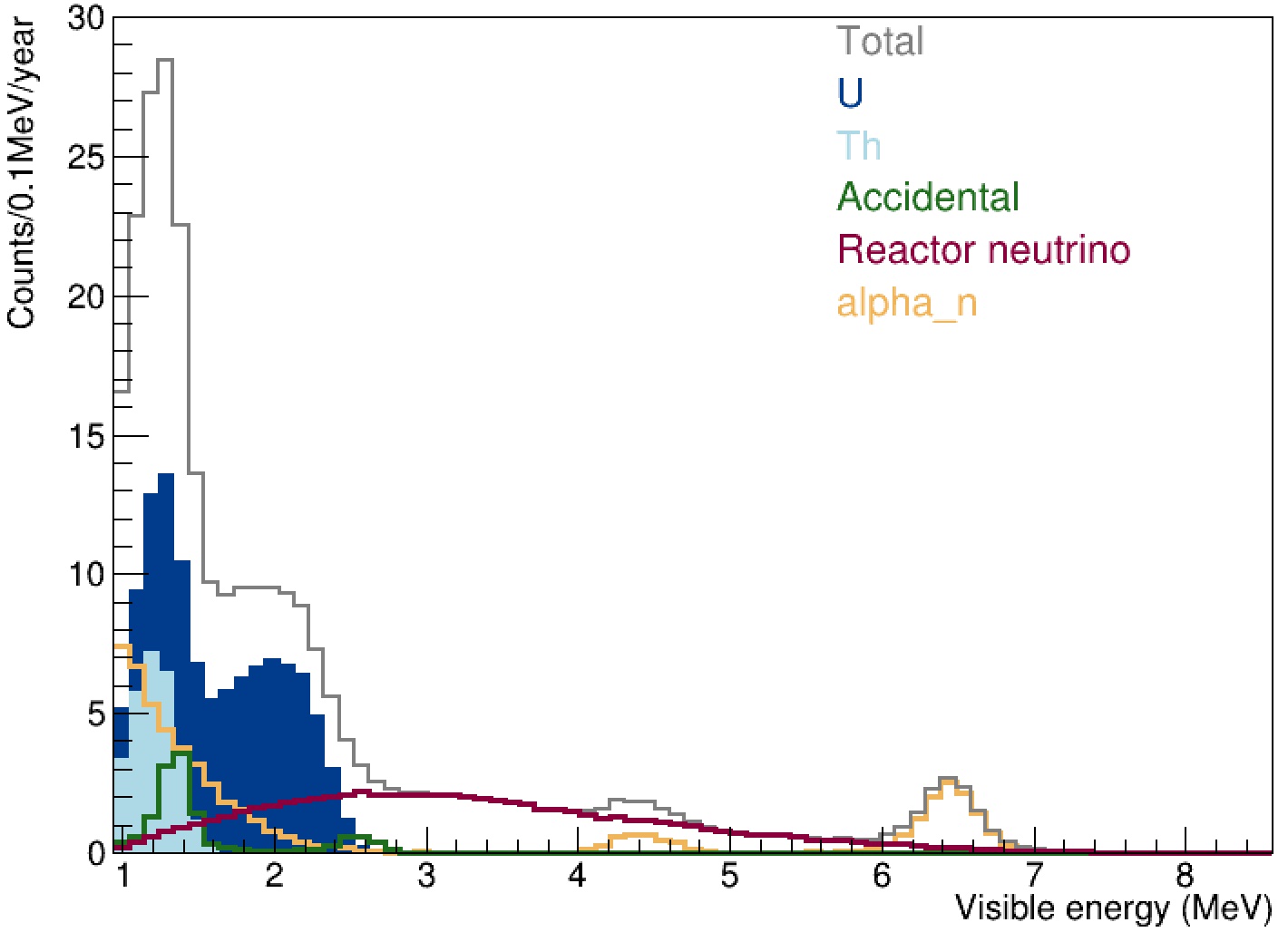}%
            }%
        }

    \put(72,67.3){%
            \setlength{\fboxsep}{2pt}%
            \colorbox{white}{%
                \parbox[t]{1.5cm}{\raggedright\tiny
                    \textcolor{gray}{Total}\\
                    \textcolor{Blue}{U}\\
                    \textcolor{Cyan}{Th}\\
                    \textcolor{Green}{Accidental}\\
                    \textcolor{Red}{Reactor}\\
                    \textcolor{Orange}{$(\alpha, n)$}%
                }%
            }%
        }
    \end{overpic}
    \caption{Spectrum of geoneutrino signal and primary background in a simulated detector geometry, with an inset plot showing the full energy range. Figure is adapted from~\cite{Ohno2024ComprehensiveSensitivityOBD}.}
    \label{fig_spectrum_BG}
\end{figure}

A detailed simulation trade study compared 1.5~kt \cite{Sakai2021DevelopmentOBD} and 17~kt designs~\cite{Ohno2024ComprehensiveSensitivityOBD}. The analysis included major background sources, including global reactor antineutrinos, accidental coincidences driven by radioactivity in detector materials, and intrinsic radiogenic and cosmogenic contributions. In particular, the work highlights $ (\alpha,n) $ production, including the well-known $ ^{13}\mathrm{C}(\alpha,n)^{16}\mathrm{O} $ channel that can be fueled by radon-chain contamination (including scenarios with $ ^{222}\mathrm{Rn} $ at 30~$ \mathrm{Bq\,m^{-3}} $). The study used realistic radiopurity assumptions and produced cost estimates, noting that a smaller detector typically requires longer operating time to reach comparable physics sensitivity, which directly affects lifecycle cost per measuring campaign. The signal and primary backgrounds  for a 17-kt detector are shown in Fig.~\ref{fig_spectrum_BG}, and the corresponding event rates  are listed in Table~\ref{tab:predicted_event_counts}. Sensitivity and operation cost estimates are listed in Table~\ref{tab:summary_sensitivity_costs}.

\begin{table*}[t]
\centering
\begin{tabular}{||l||cc|c||cccc|c||}
\hline\hline
  & \multicolumn{3}{c||}{Geoneutrinos} & \multicolumn{5}{c||}{Backgrounds} \\
\cline{2-4}\cline{5-9}
 & U & Th & Total & Reactor & Accidental & ($\alpha$,n) & He-Li & Total \\\hline\hline
Full energy region
  & \multirow{2}{*}{100}
  & \multirow{2}{*}{25}
  & \multirow{2}{*}{125}
  & 68 & 14 & 55 & 0 & 138 \\
Geoneutrino region
  &  &  &
  & 23 & 13 & 40 & 0 & 76 \\\hline
Mantle contribution
  & 75 & 19 & 94
  &  ---    &  ---    &  ---    & --- &  ---     \\
\hline
\end{tabular}
\caption{Predicted annual event counts for a 17-kt detector. For geoneutrino signal, a 12.1-TNU  value was used, corresponding to the Middle-Q model. Cosmogenic $\beta$-n emitters such as $^{8}\mathrm{He}$ and $^{9}\mathrm{Li}$ were also included but found to be negligible~\cite{Ohno2024ComprehensiveSensitivityOBD}. }
\label{tab:predicted_event_counts}
\end{table*}

\begin{table*}[htbp]
  \centering
  \resizebox{\textwidth}{!}{%
  \begin{tabular}{|l||cc|cc|cc||ccc||}
    \hline
    & \multicolumn{6}{c||}{Zero mantle-signal exclusion} & \multicolumn{3}{c||}{Costs (in oku yen)} \\
    \cline{2-7}\cline{8-10}
    Detector &
    LowQ $\sigma$ & LowQ period &
    MiddleQ $\sigma$ & MiddleQ period &
    HighQ $\sigma$ & HighQ period &
    Fabrication & Ship construction & Ship operating \\
    \hline
    1.5 kt &
    2.5$\sigma$ & 10 years &
    3.5$\sigma$ & 3 years &
    3.7$\sigma$ & 1 year &
    48 & 28 & 21 $\times$ 10 years \\
    17 kt  &
    3.6$\sigma$ & 1 year &
    4.2$\sigma$ & 3 months &
    4.3$\sigma$ & 1 month &
    369 & 78 & 57 $\times$ 1 year \\
    \hline
  \end{tabular}%
  }
  \caption{Summary of significance, observation period, and cost by detector size~\cite{Ohno2024ComprehensiveSensitivityOBD}. Oku yen is 100 million yen and roughly corresponds to 1 million U.S. dollars. Low, middle, and high Q values correspond to values listed in Table 1.}%
  \label{tab:summary_sensitivity_costs}
\end{table*}

\subsubsection{Liquid scintillator performance at deep ocean temperature}
\begin{figure}[h]
     \centering
     \includegraphics[width=\linewidth]{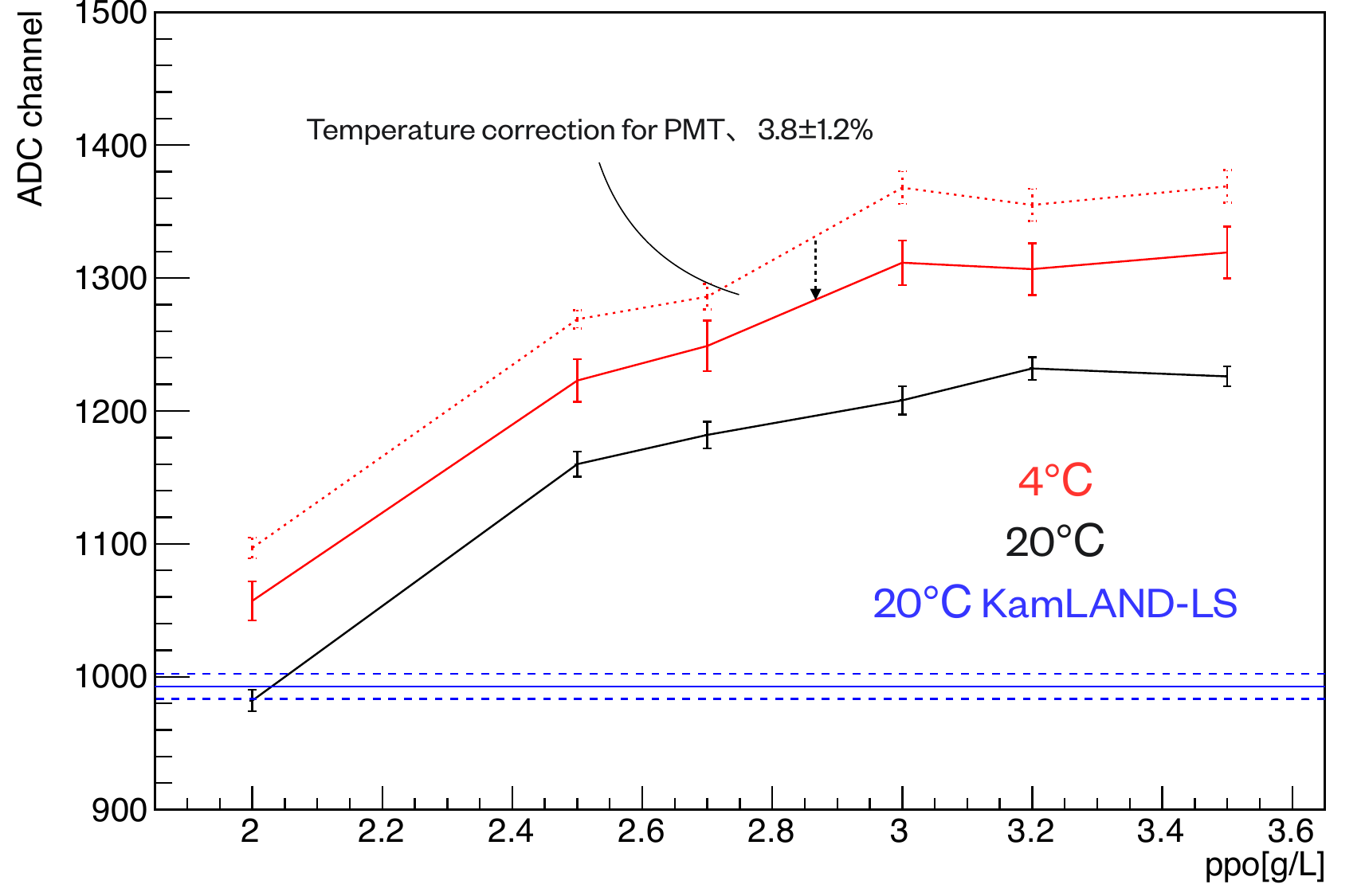}
     \caption{Measured light yield as a function of PPO concentration in the LAB-based liquid scintillator. The vertical axis (ADC channel) corresponds to the energy deposited by \({}^{137}\)Cs \(\gamma \)-rays detected by the PMT. The blue line represents the KamLAND LS at 20\({}^{\circ }\mathrm{C}\). The black and red lines represent the LAB-based LS at 20\({}^{\circ }\mathrm{C}\) and 4\({}^{\circ }\mathrm{C}\), respectively. The gap between the red dotted and solid lines corresponds to the PMT temperature correction required due to the drift in PMT quantum efficiency.}
     \label{fig_LS}
\end{figure}

A dedicated low-temperature performance study was conducted for a laboratory-based liquid scintillator pertinent to OBD~\cite{Sakai2021DevelopmentOBD}. 
The liquid scintillator consists of LAB (Linear Alkyl Benzene) and PPO (2,5-diphenyloxazole) as a fluor. The temperature of the liquid scintillator, contained within an airtight 100~ml stainless steel container, was regulated at either 20$^\circ\mathrm{C}$ or 4$^\circ\mathrm{C}$, with the latter selected to replicate deep-ocean conditions. 
Scintillation light was detected via a 2-inch photomultiplier tube positioned beneath the vessel's transparent viewing window. 
Using Compton scattering from a $^{137}$Cs gamma source, the study evaluated the scintillation light output as a function of PPO concentration. 
The findings revealed that the light yield at 4$^\circ\mathrm{C}$ is systematically higher than at 20$^\circ\mathrm{C}$ across the entire tested concentration range. 
As clearly illustrated by the trend in Figure~\ref{fig_LS}, cooling the scintillator to deep-ocean temperatures consistently enhances the light yield.

For the candidate OBD formulation near a PPO concentration of 3.0~g/L, the study reports an approximately 8.6\% increase in scintillation response at 4$^\circ\mathrm{C}$ relative to 20$^\circ\mathrm{C}$, after accounting for the independently measured PMT temperature effect. 
The PMT contribution alone was determined to be 3.8 $\pm$ 1.2\%, implying that the remaining gain is directly attributable to the scintillator itself. 
The study also showed that LS transmittance does not vary
significantly with temperature.
In practical terms, this result is encouraging for ocean-bottom deployment; it demonstrates that the LAB+PPO scintillator retains excellent optical performance at deep-sea temperatures and provides modestly improved photon statistics, thereby supporting low-energy event detection without requiring a fundamentally alternative scintillator formulation.

\begin{figure}[htbp]
    \begin{overpic}[width=1\linewidth]{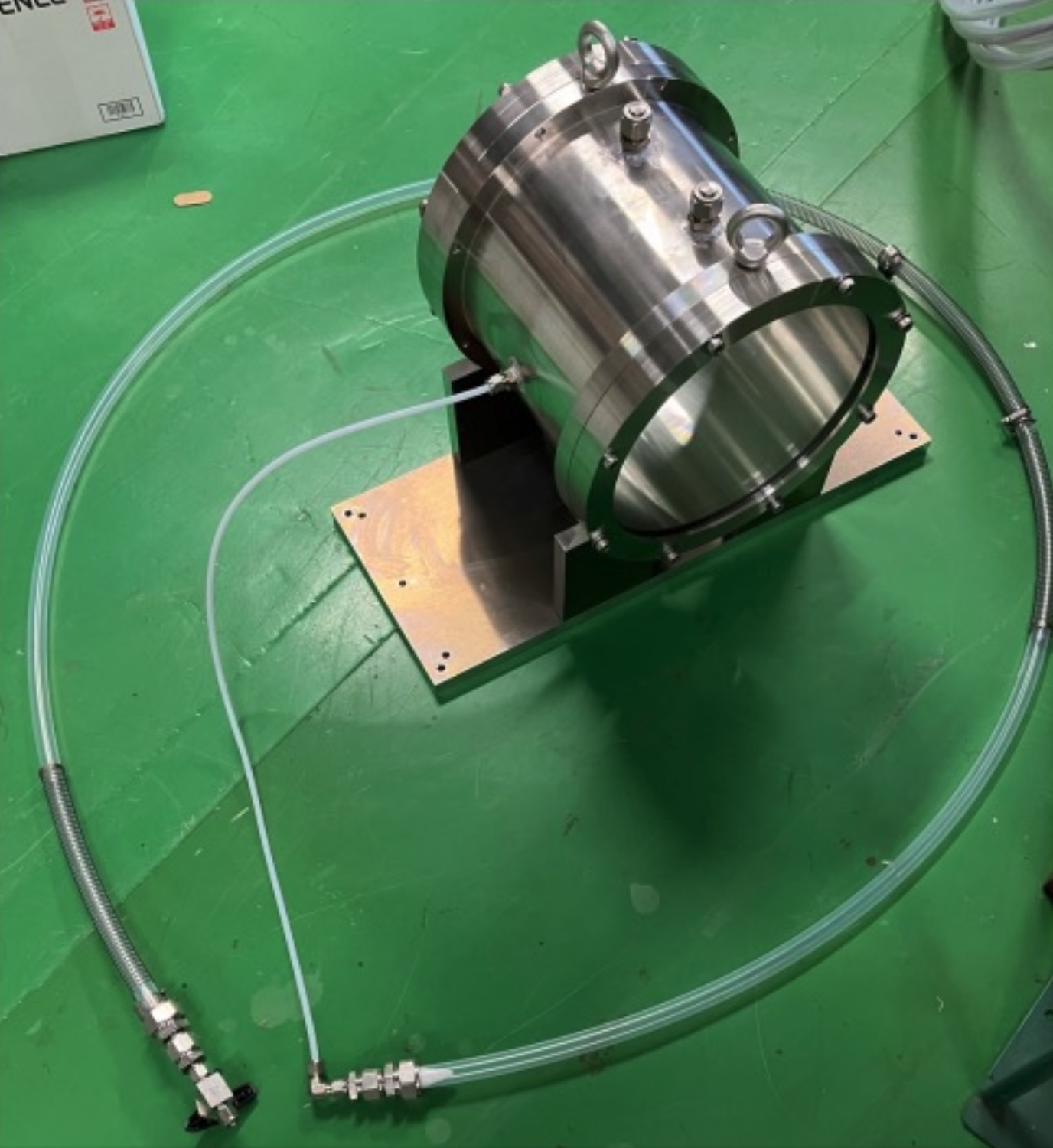}
    \put(25,90){\color{white}\vector(1,0){30}}
    \put(20,90){\fcolorbox{red}{white!40}{\parbox{30pt}{\centering \textcolor{black}LS fill ports}}}
    \put(15,67){\color{white}\vector(1,0){28}}
    \put(5,67){\fcolorbox{red}{white!40}{\parbox{60pt}{\centering \textcolor{black}Pressure compensation tube}}}
    \put(60,35){\color{white}\vector(0,1){20}}
    \put(50,35){\fcolorbox{red}{white!40}{\parbox{40pt}{\centering \textcolor{black}Acrylic window}}}
    \end{overpic}
\\\hspace{0.5cm}
    \begin{overpic}[width=1\linewidth]{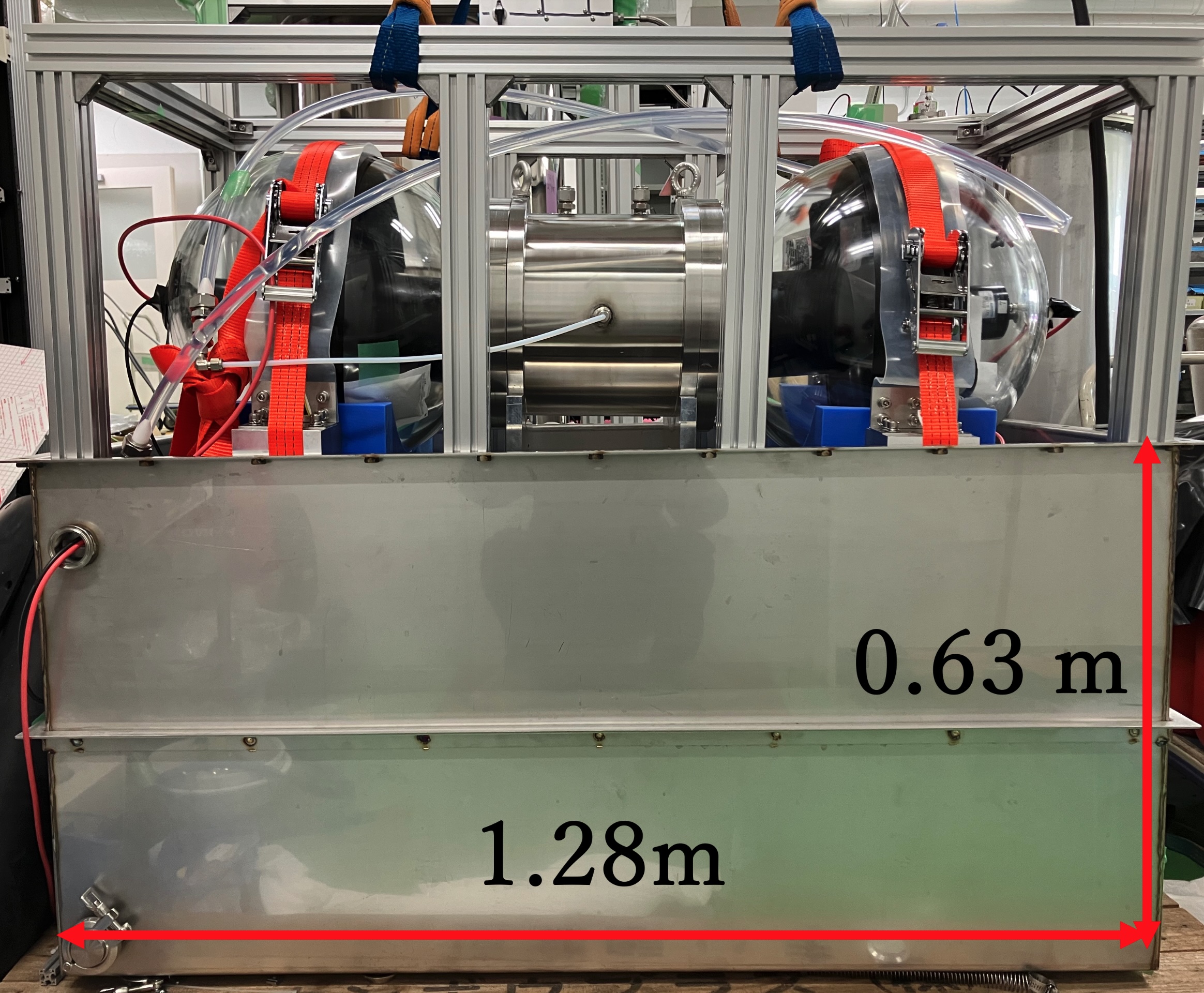}
    \put(50,35){\color{white}\vector(-1,1){20}}
    \put(50,35){\color{white}\vector(1,1){20}}
    \put(35,35){\fcolorbox{red}{white!40}{\parbox{60pt}{\centering \textcolor{black}PMTs inside glass spheres}}}
    \end{overpic}    
    
    \caption{A 12-liter prototype constructed at Tohoku University to study performance of the liquid scintillator at the ocean-bottom conditions --- the cylindrical stainless steel body with two acrylic windows for PMT attachment. The photograph below shows fully instrumented prototype ready for deployment: the PMTs are inside glass spheres to sustain high pressure.}
    \label{fig_12l_proto}
\end{figure}

\subsubsection{Optical performance of a prototype detector} %

A prototype detector constructed by Tohoku University (shown in Fig.~\ref{fig_12l_proto}) was used to determine its performance in a marine environment \cite{Chauhan2025OpticalPerformance}. This prototype consisted of a cylindrical stainless steel vessel, with an outer radius of 14 cm, an inner radius of 13 cm, and a length of 20 cm, with acrylic windows bolted to both ends of the vessel. 5” PMTs were positioned against the windows to collect scintillation photons from the LAB liquid organic scintillator that filled the vessel. Using a $ ^{137}\mathrm{Cs} $ gamma source placed just outside the stainless steel wall of the prototype, as well as simulations performed in Geant4, the light collection efficiency and intrinsic resolution of the scintillator and prototype were calculated for both surface conditions and marine conditions. The intrinsic resolution was found to improve significantly for marine conditions in comparison to surface conditions. The accuracy of the simulations performed was also evaluated and found to be rather accurate to the detector measurements for surface conditions. The accuracy of the marine conditions simulation has yet to be evaluated, with the next step being to deploy the prototype detector underwater~\cite{Chauhan2025OpticalPerformance}.

\subsubsection{Structural design and selection of buffer oil and acrylic vessel} %
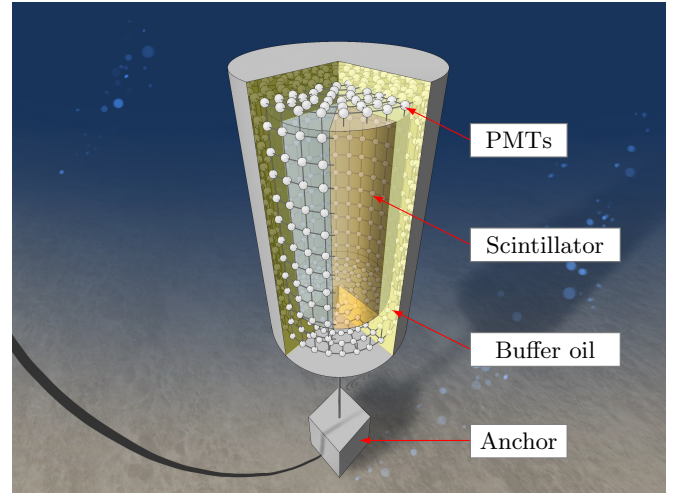
\begin{figure}[htbp]
    \centering
    \begin{overpic}[width=1\linewidth]{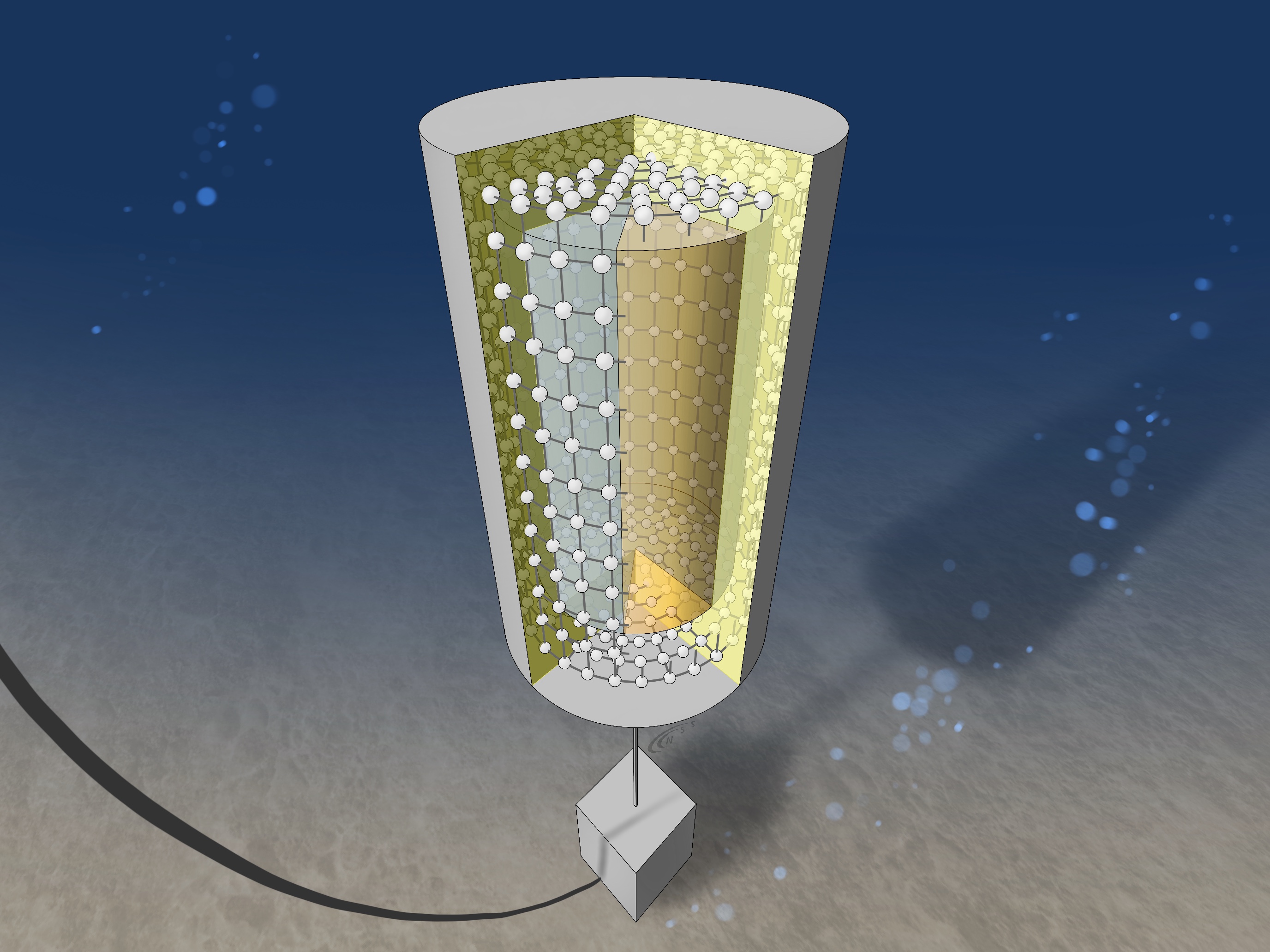}
    \put(70,54){\color{red}\vector(-2,1){10}}
    \put(70,53){\fcolorbox{gray}{white!40}{\parbox{30pt}{\centering \textcolor{black}PMTs}}}
    \put(70,22){\color{red}\vector(-2,1){12}}
    \put(70,21){\fcolorbox{gray}{white!40}{\parbox{50pt}{\centering \textcolor{black}Buffer oil}}}
    \put(70,38){\color{red}\vector(-2,1){15}}
    \put(70,37){\fcolorbox{gray}{white!40}{\parbox{50pt}{\centering \textcolor{black}Scintillator}}}
    \put(80,8){\color{red}\vector(-1,0){27}}
    \put(70,7){\fcolorbox{gray}{white!40}{\parbox{30pt}{\centering \textcolor{black}Anchor}}}
    \end{overpic}
    \caption{Conceptual diagram of the ocean-bottom neutrino detector. An inner acrylic vessel contains liquid scintillator, surrounded by buffer oil between the acrylic and inward-facing photosensors.} %
    \label{fig:detector}
\end{figure}

A conventional detector concept is shown in Fig.~\ref{fig:detector}. As in land-based liquid scintillator detectors (e.g., KamLAND, SNO+, JUNO), an inner acrylic vessel contains liquid scintillator (LS) and is surrounded by a buffer region instrumented with inward-facing PMTs. Outward-facing PMTs provide a muon veto, with the surrounding ocean water acting as the veto and shielding volume.

Buffer oil (BO) serves to attenuate backgrounds from radioactivity in PMTs and nearby structures, and to provide optical coupling between the LS volume and the inward-facing PMTs. While the LS solvent was selected as LAB, the BO composition required dedicated evaluation. BO should have high optical transmittance, a refractive index similar to the LS to minimize optical losses, and a density similar to the LS to reduce differential loads on the acrylic vessel. BO candidates were screened for transparency and density, then characterized via transmittance, refractive index, and density measurements~\cite{Ono2026DetectorStructureOBD}.

The acrylic vessel holds LS internally and BO externally, and bears sustained loads due to any density difference between the two liquids. Polymeric materials such as acrylic can gradually deform under sustained stress (creep), potentially reducing long-term structural margin. Based on the anticipated deployment lifetime ($\sim$10 years) and measured creep behavior, an allowable stress criterion for candidate acrylics was established.

The structural design established two requirements. First, the detector must operate safely and without damage over long-term measurements ($\sim$10 years) under deep-sea conditions (low temperature and high hydrostatic pressure). Second, the design should maximize light collection efficiency, which governs energy and position resolution. The target light collection efficiency was set to 450\,p.e./MeV, comparable to KamLAND \cite{inoue2004reactor}, consistent with prior simulations demonstrating feasibility for mantle geoneutrino observations and discrimination among radiogenic heat models.

\begin{figure}[htbp]
    \centering
    \includegraphics[width=1.\linewidth]{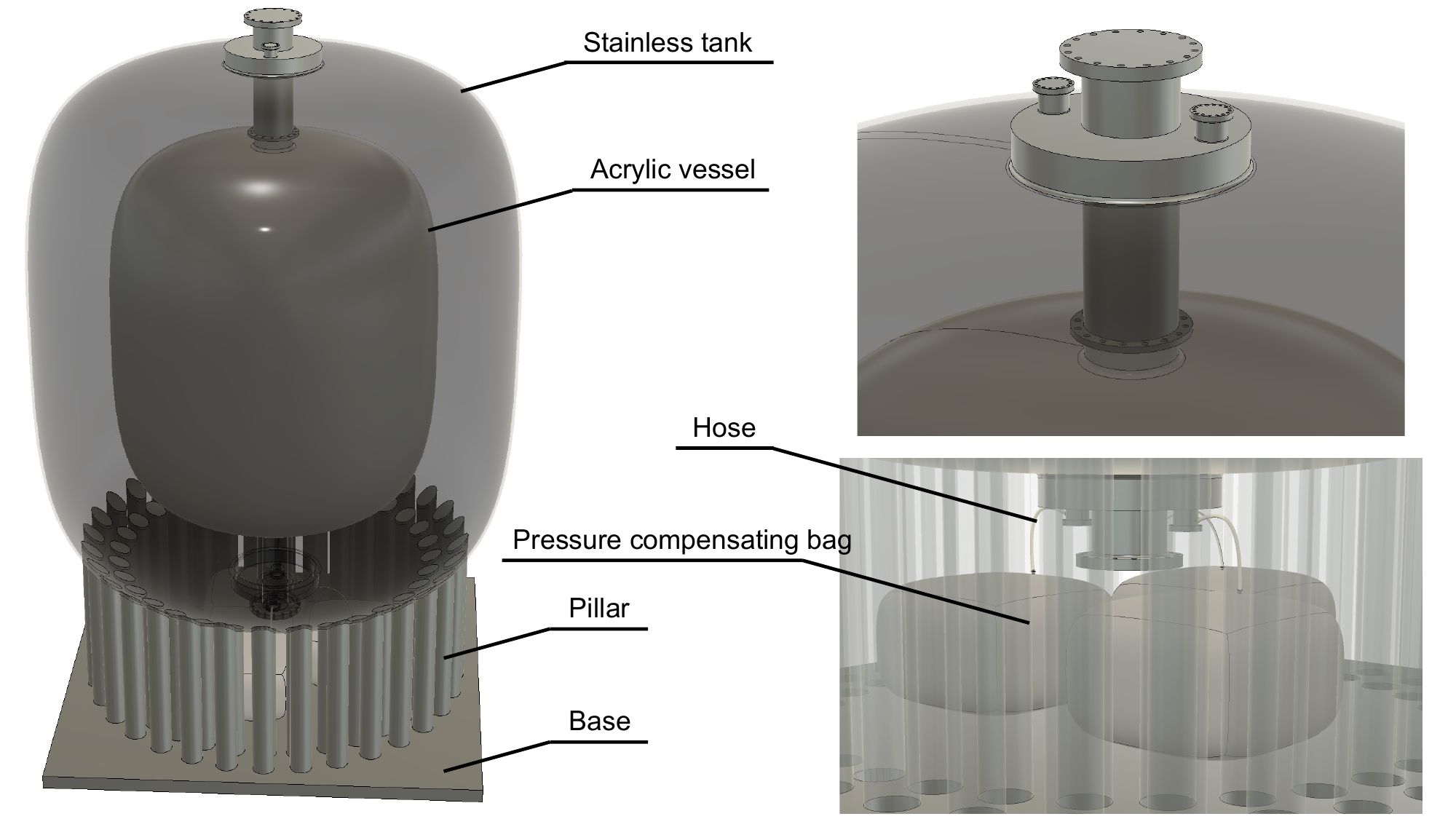}
    \caption{Baseline geometry used for structural analysis.}
    \label{fig:geometry}
\end{figure}

The study examined the 1.5-kt detector geometry in Fig.~\ref{fig:geometry}. Two design choices were adopted: (i) cylindrical stainless-steel tank and acrylic vessel to maximize LS volume under ship transport constraints, and (ii) a BO thickness of 3\,m, previously optimized to balance accidental backgrounds and light collection. The resulting dimensions were a stainless-steel tank of radius 9\,m and height 20\,m, and an acrylic vessel of radius 6\,m and height 14\,m. However, component thicknesses remain a significant optimization space affecting fabrication cost, optical performance, and safety, particularly because the OBD is cylindrical (not spherical) and must withstand deep-ocean pressure ($\sim$40~MPa, at 4-km depth).

The design procedure was as follows. The detector geometry (stainless-steel tank, acrylic vessel, pressure compensation mechanism, support pillars, and seabed foundation) was modeled for stress analysis. Detailed modeling of LS, BO, and photosensor internals was deferred due to the complexity of fluid-structure coupling. Instead, loads were represented via hydrostatic pressures of LS, BO, and seawater, together with weight and buoyancy of the photosensor modules. Stress analyses were repeated while varying acrylic and stainless-steel thicknesses. For stainless steel, a safety factor $\geq 2$ was taken as sufficient. For acrylic, a maximum stress $\leq 9.6$\,MPa was adopted, based on creep test results with a safety factor of 2 for $\sim$10 years of operation. Under this allowable, the estimated creep life is $\sim$56 years. Among candidate structures meeting these requirements, the design maximizing light collection efficiency was selected as the baseline.

\section{Direction-sensitive Geoneutrino Detection}

\subsection{Directionality via a pattern-matching algorithm}
The detection of geoneutrinos has been based on the IBD reaction \eqref{IBD}.
In this reaction, the outgoing neutron retains a correlation with the direction of the incident antineutrino.
Due to the reaction kinematics, the neutron is emitted preferentially in the forward direction, such that the displacement between the positron interaction point (prompt signal) and the subsequent neutron-capture location (delayed signal) can, in principle, be used to infer the incoming antineutrino direction \citep{vogel1999angular}.
However, in standard liquid scintillators used in both existing and decommissioned geoneutrino detectors, the neutron undergoes a prolonged random walk before capture, which washes out this initial displacement and destroys the directional information \citep{tanaka20146li}.

In practice, this directional information can be enhanced by doping the scintillator with nuclei that have large neutron-capture cross sections, such as $^{6}\mathrm{Li}$ or Gd, which substantially suppress neutron diffusion before capture \citep{tanaka20146li,seo2020labls}.
Such directionality has already been demonstrated in several reactor neutrino experiments \cite{hochmuth2007exploiting,andriamirado2025reactor}.
Recently, an algorithm has been developed to enhance the angular sensitivity of neutrino detectors. The algorithm uses pattern matching to identify the source direction and has a theoretical basis for the fit it uses \cite{10.1063/5.0315079}. Although it was developed in the context of determining source direction for reactor antineutrinos and for segmented detectors, it can in principle be used for geoneutrino since the underlying detection mechanism is the same. The algorithm was also found to have a reasonable angular sensitivity even at low event counts \cite{crow2026enhancingangularsensitivitysegmented}. This will enable the detector to identify distributed sources.

\subsection{Measurability of HPE abundances in LLVPs}

Some sensitivity studies based on a 1.5-kt OBD configuration demonstrated the feasibility of measuring HPE abundances in LLVP regions via directional geoneutrino detection \cite{Xu2026hhx,xu2026proceedings}. In these works, the mantle outside the LLVPs was assigned HPE abundances typical of a depleted MORB-source reservoir \cite{arevalo2010chemical}, while the remaining mantle HPE inventory was concentrated entirely within the LLVP regions. The results showed that direction-sensitive detection at Pacific Ocean sites not only enables the discrimination of LLVPs, but also permits the inversion of their internal HPE concentrations, thereby constraining their radiogenic heat production.

Using the same Earth models as in \cite{xu2026proceedings}, we evaluate the sensitivity to LLVP structures using a profile likelihood ratio test.
Sensitivities are evaluated for both 1.5-kt and 17-kt OBD configurations, extending previous studies that focused exclusively on the 1.5-kt case.
To calculate the median expected significance, we adopt the Asimov dataset approach based on the asymptotic formulae from \cite{cowan2011asymptotic}.
The null hypothesis ($H_0$) assumes a uniform radioactive distribution in the mantle, while the alternative hypothesis ($H_1$) includes LLVP structures enriched in HPEs.
For the heterogeneous scenario, two LLVP-like regions are modeled immediately above the core-mantle boundary (CMB): one beneath the Pacific Ocean with a radius of 35$^\circ$ and one beneath Africa with a radius of 40$^\circ$, both extending up to 1000 km in height and located at nearly antipodal positions.
In both the homogeneous and heterogeneous cases, the crust is modeled using CRUST1.0 \cite{laske2013update} to ensure a realistic treatment of the local geological structure.
We consider both 1.5-kt and 17-kt OBD configurations and include realistic backgrounds such as reactor antineutrinos and detector backgrounds as reported in \cite{Sakai2021DevelopmentOBD,Ohno2024ComprehensiveSensitivityOBD}.
Specifically, the detector backgrounds are set equivalent to KamLAND's initial high-noise phase \cite{kamland2015solarnu} to ensure a conservative assessment.

\begin{figure}[htbp]
    \centering
    \includegraphics[width=1.\linewidth]{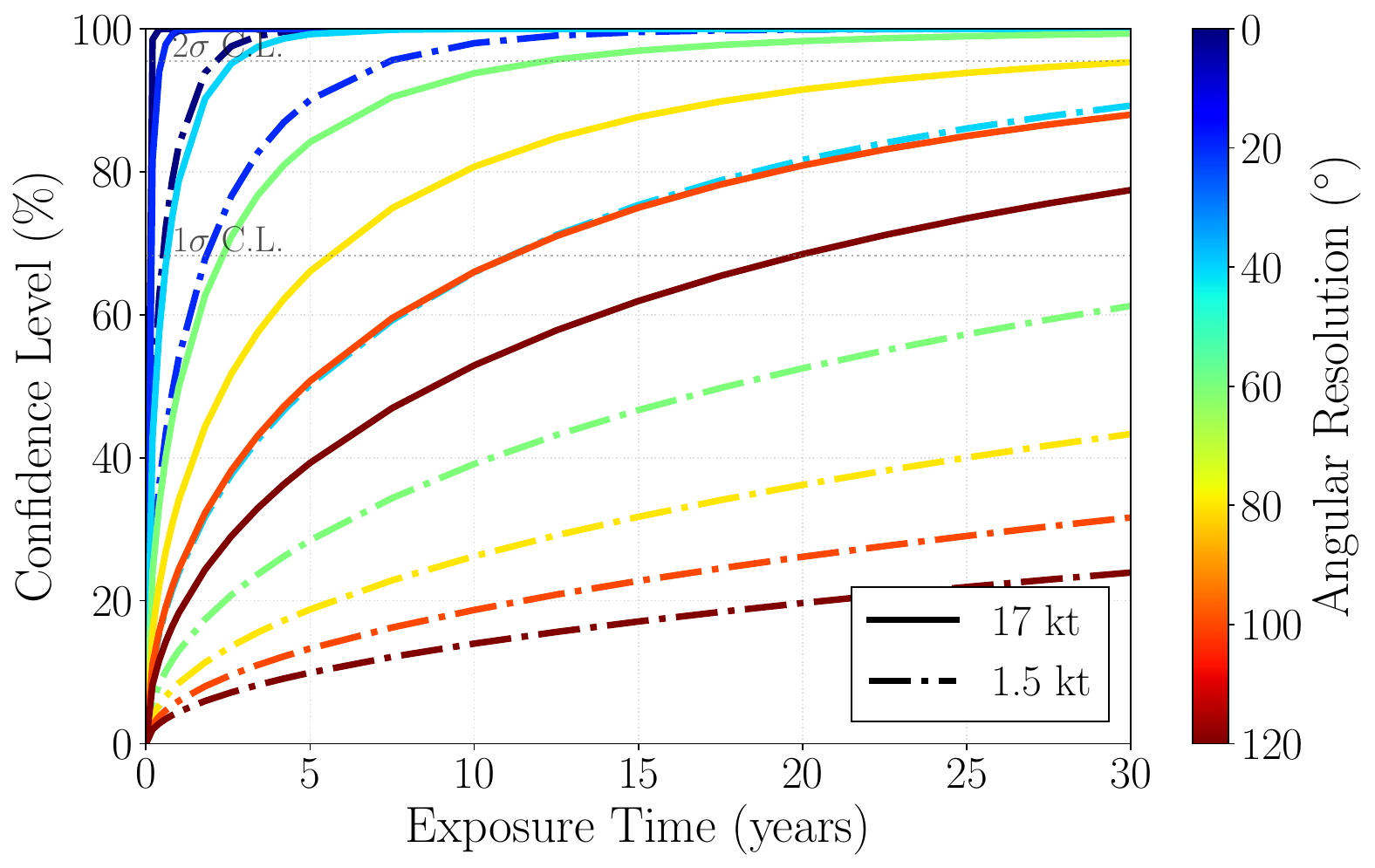}
    \caption{Expected significance for rejecting the homogeneous mantle hypothesis ($H_0$) as a function of exposure time and detector angular resolution. The dash-dotted and solid lines represent the 1.5-kt and 17-kt OBD configurations, respectively, with colors mapping different angular resolutions.}
    \label{fig:LLVPsensitivity}
\end{figure}

Figure \ref{fig:LLVPsensitivity} shows the expected exclusion significance of the homogeneous mantle hypothesis ($H_0$) hypothesis as a function of detector exposure and angular resolution.
The results for the 1.5-kt and 17-kt OBDs are depicted by dash-dotted and solid lines, respectively, with colors indexing different angular resolutions.
The results indicate that finite angular resolution significantly improves the ability of a single detector to identify LLVP structures, which in turn enables studies of HPE concentrations within these structures.
Also, the larger target mass of the 17-kt configuration substantially increases the geoneutrino event rate, thereby reducing the exposure time required to accumulate sufficient statistics and allowing a high confidence level to be achieved within the initial years of operation.

\section{Impact beyond geoscience}

\subsection{Supernova neutrino observation}
In addition to detecting neutrinos originating from within the Earth, the proposed OBD detector would also be sensitive to neutrinos from a Galactic core-collapse supernova (CCSN) event, thus proving key insights into the supernova mechanism. Current 3D radiation-hydrodynamic CCSN simulations are able to template predicted neutrino signatures \cite{2024ApJ...964L..16B,2025ARNPS..75..425J,2018ApJ...865...81O,2013ApJ...766...43M,2012ApJ...755...11K}.%
The neutrino event rate for an OBD-like detector, assuming a supernova occurring within a distance of 10 kiloparsecs, would be of the order of several hundred neutrinos per second, surpassing the $\sim$19 neutrinos detected from SN1987A, the only supernova neutrino signal observed to date \cite{hirata1987observation,bionta1987observation,Alekseev1988}. Such a high event rate would enable us to follow the stages of the core-collapse supernova process in real time, including a gradual rise to a plateau as material from the outer layers of the star accretes onto the proto-neutron star forming in the core, followed by a decay in neutrino emission as the star enters the Kelvin-Helmholtz cooling phase. An abrupt cutoff to the neutrino event rate seen by the detector would provide a clear signature of black hole formation. 

Neutrino detection would also allow rare insight into the stellar core structure in the moments of core collapse, revealing the otherwise opaque core density profile and constraining the tenuously understood nuclear equation of state \cite{2025PhRvD.111l3038C,2019MNRAS.489.2227V,2023MNRAS.526.5900V,2023PhRvD.108f3020N,2022MNRAS.512.2806N,2021MNRAS.506.1462N,2021MNRAS.500..696N,2018MNRAS.480.4710S,2016ApJ...817..182W, 2025MNRAS.540.3885B,2026ApJ..1003...40R}. Stars with different core compactness are predicted to differ by several hundred events per second in peak neutrino rate, and the high event rate in the detector would enable discrimination between stellar progenitor models. Theoretical correlations between the magnitude of the peak luminosity seen by the detector and remnant properties may also allow for the inference of the mass of the residual neutron star formed by the event. Finally, the amplitude of the signal, which is sensitive to neutrino oscillation physics, may place constraints on the neutrino oscillation model and mass hierarchy. Neutrino-driven instabilities in the first seconds of CCSN will shape the late-time (day-to-years) morphology of the ejecta distribution \cite{2025ApJ...982....9V, 2025arXiv250916314V,gabler}.

This OBD would complement the detections of other existing and upcoming detectors such as Super-Kamiokande, DUNE, and JUNO when a Galactic supernova occurs. As there are uncertainties in and degeneracies between neutrino luminosity, oscillation model, and progenitor compactness, the detection of a supernova neutrino lightcurve within multiple detectors is essential to accurately characterizing and disentangling the physical information encoded in the neutrino signal. 

The OBD will also be deployed in the southern hemisphere during its lifetime. This will be the only antineutrino detector capable of detecting SN neutrinos and geoneutrinos in the southern hemisphere.

\subsection{High energy physics}

The ability to vary the baseline over a wide range of distances in the far field (tens of kilometers) offers a unique opportunity to systematically map the oscillation probability space, a capability that has not yet been realized in existing experiments. 

Current reactor experiments such as KamLAND, along with its ongoing KamLAND2 upgrade, face a "blurry" baseline, as the effective distance is averaged over contributions from multiple reactors, limiting precision in the measurement of oscillation parameters \cite{gando2013reactor}.
A setup involving a single, isolated power reactor located near a shoreline, combined with measurements taken at various baselines, would provide a more controlled environment to probe fundamental neutrino oscillation physics. Such an experimental configuration could also offer sensitivity to physics beyond the Standard Model (SM).
While the JUNO experiment, which has come online, will make significant contributions to the field, other detectors such as SNO+ face statistical limitations due to the large distances separating the reactor sources from the detector. Furthermore, none of the existing far-field land-based detectors are equipped to vary baselines, which constrains their ability to explore oscillation phenomena across a broader parameter space, without relying on the energy binning to effectively sample various $L/E$ values.

The OBD can in principle serve as a far detector in the long-baseline accelerator-neutrino context. The central advantage of a mobile deep-ocean far detector is the ability to choose and adjust the baseline (like in the reactor context) and off axis angle after the beam has been constructed. For fixed underground detectors, the baseline and off-axis angle are effectively locked in by geography and civil engineering constraints. In contrast, a relocatable deep ocean detector can be deployed at a range of baselines along the beam axis, and at selected off axis angles defined by available deep water locations, allowing detailed optimization of oscillation physics reach.

Baseline flexibility allows one to probe different oscillation maxima and matter effect regimes using the same beam and detector technology. For example, a Tokai-to-Kamioka (T2K) beam \cite{abe2011t2k} aimed toward a deep ocean site at longer baseline (like in the proposed Tokai-to-Kamioka-Korea experiment \cite{hyper2018physics}) could enhance sensitivity to mass ordering and subdominant CP violating effects by exploiting stronger matter effects over the longer path length. A mobile far detector could first be operated at one baseline to collect an initial data set, then redeployed to a second baseline to test degeneracies and cross check oscillation parameter fits with different $L/E$ coverage, without constructing a new cavern. This approach mirrors, at far detector scale, the “spectral shaping” concept that motivates a vertically moving detector, such as Intermediate Water Cherenkov Detector (IWCD) at the near site \cite{scott2016intermediate,drakopoulou2018intermediate}, but now uses geographic relocation to tune the oscillation pattern.

Off-axis angle control provides additional leverage. As in the IWCD concept, the accelerator neutrino energy spectrum depends strongly on off-axis angle. While a far detector cannot scan angles continuously in the way a small near detector can, a mobile platform can be positioned at a discrete set of deep ocean sites that correspond to different off-axis angles, each providing a different beam energy spectrum. In principle, a sequence of deployments at selected angles could mimic narrower or broader spectra and provide complementary sensitivity to appearance and disappearance channels, or optimize the balance between event rate and energy resolution for CP violation studies.

A mobile deep-ocean detector can, beyond its scientific mission, drive next generation HEP technology R\&D because it must operate at large scale with tight power and bandwidth constraints and only remote servicing. Those requirements motivate development of marine compatible scintillating media (and potentially enviromentally friendly detection media), including hybrid water-based scintillators \cite{yeh2011new,land2021mev,Ford:2022wla}, to improve low energy performance and particle identification; large area photosensors such as advanced SiPM arrays \cite{piemonte2019overview} and LAPPDs \cite{adams2016brief} to improve photon detection efficiency, timing, and spatial resolution relative to PMTs; highly integrated low power multi channel SoC readout electronics designed for pressure tolerant deep sea housings; real time data reduction and intelligent triggering using AI/ML on FPGAs and SoCs \cite{gonski2026AI} to compress data in situ, identify rare signatures, and adapt to changing source and background conditions; and improved reconstruction and analysis algorithms, including ML-based event classification and energy reconstruction \cite{li2023kamnet}. Many of these developments are of relevance to future HEP detectors, such as DUNE \cite{falcone2022dune}. It also drives innovation in high-pressure enclosures, connectors, feedthroughs, and electronics packaging relevant to neutrino telescopes, such as IceCube \cite{halzen2010icecube} and KM3NeT \cite{margiotta2014km3net}. Because the detector is modular and relocatable, these technologies can be deployed, upgraded, and benchmarked across multiple campaigns, accelerating maturation for future underground and underwater detectors, long-baseline experiments, and other low power, high granularity autonomous sensing applications in harsh environments.

\subsection{Nonproliferation and nuclear reactor monitoring capabilities}

Although the primary motivation for an OBD is fundamental neutrino science, the system naturally offers a set of attractive capabilities for nuclear monitoring and nonproliferation applications. The key distinguishing feature is mobility: unlike large underground detectors, the OBD can be engineered as a relocatable asset that is deployed and recovered using standard marine operations. Once the detector technology and support infrastructure are in hand, this provides a standing capability that can be moved on timescales of months to specific regions of interest, including areas where terrestrial access is politically or logistically difficult. In practice, a small number of such instruments could, over time, survey an extensive geographic region by redeployment rather than new construction at each site.

From an infrastructure perspective, ocean bottom deployment avoids the need for excavation or the creation of large underground cavities, which has been a dominant cost and schedule driver for land-based detectors. The ocean water column provides a natural, uniform overburden, simplifying shielding requirements and decoupling the detector cost from site specific civil engineering. This architecture is inherently modular and extensible; additional scientific or application specific payloads, including seismometers, oceanographic sensors, or dedicated monitoring instruments, can be integrated into the same platform. As new monitoring modalities or analysis techniques are developed in the future, they can be incorporated into subsequent OBD deployments without re‑engineering a fixed underground site.

In the nonproliferation context, OBD is particularly relevant for monitoring coastal nuclear facilities and future ocean-based energy systems, such as floating nuclear power plants or maritime reactor platforms, for which an ocean coastline is effectively “closer” than any land border. Previous analyses have shown that allowing detector deployment at sea, at coastline rather than traditional land border distances, increases the number of facilities that are observable with finite detector masses~\cite{Cogswell01092024}. At the 12 nautical mile territorial sea limit (about 22~km), even a relatively small 50-MW reactor produces of order 100 IBD events per kiloton per year, which is a meaningful signal level for long term monitoring. The OBD capability positioned just outside territorial waters can therefore provide a unique, non‑intrusive complement to traditional monitoring tools in selected cases, especially where direct access on land is constrained.

At the same time, there are important limitations that shape realistic monitoring use cases. If the detector is optimized primarily for reactor monitoring, it will generally operate at depths and locations where reactor antineutrino flux dominates, which significantly complicates precision measurements of geoneutrinos. Moreover, when deployed near a continental margin, the local crustal geoneutrino flux dominates and the mantle contribution remains subdominant, which further limits sensitivity to global mantle properties \cite{huang2013reference}. Depth requirements for physics performance and background reduction also impose strong constraints on where the OBD can be deployed. Finally, the relevant legal and diplomatic frameworks, including the 12 nautical mile territorial sea definition and broader issues of operating scientific sensors near another state’s critical infrastructure, will constrain where and how such a system can be used. Taken together, these considerations imply that the OBD should be viewed not as a universal monitoring solution, but as a flexible, relocatable asset that can provide targeted, high-value neutrino-based information for a subset of coastal and maritime nuclear systems, as a “bonus” capability layered on top of their core scientific mission.

\section{Conclusion}
Up to this point, the understanding of Earth's geochemical composition has relied entirely on surface sampling, seismology, and numerical simulations. Geoneutrinos change this. For the first time, real-time observation of the Earth's interior chemical composition is possible by a method other than seismology. Neutrinos have become a mature probe of the Universe, from solar neutrinos to high-energy neutrino astronomy. By comparison, the Earth remains incompletely characterized in one of its most basic properties. These properties include the geochemical magnitude and distribution of radiogenic heating, which power thermal evolution, mantle convection, and plate tectonics.  Since the first detection of geoneutrinos by KamLAND, with subsequent measurements by Borexino, SNO+, and the upcoming JUNO experiments, neutrino geoscience has demonstrated a direct, quantitative sensitivity to heat-producing elements. At the same time, present measurements are fundamentally limited by the continental crust contribution and its uncertainties. Currently, the mantle and bulk silicate Earth radiogenic power remain poorly determined, with published measurements spanning a wide range.

A deep-ocean, relocatable antineutrino detector provides a uniquely enabling next step. 
We can position the OBD in the middle of the Pacific, and thus sampling only the mantle, from about 3000 km away from the core and continents.
Deployed far from continents, the measurement becomes mantle-dominated (the crust contribution becomes about 50--100 times suppressed compared to land-based detectors, substantially reducing an otherwise dominant crustal contribution that limits land-based experiments. 
This also means that this experiment will be far from nuclear reactors. Additionally, the deep-ocean environment offers strong shielding from cosmic-ray muons. Taken together this deployment choice provides superior background suppression. 
In this configuration, geoneutrino detection can evolve from demonstrating feasibility to delivering a definitive mantle measurement capable of discriminating competing models for Earth’s composition and heat budget.

In the past two decades, the field has advanced through improved geoneutrino statistics, multi-site measurement, and analysis maturity. In parallel, enabling technologies have progressed, including modern photosensors and low-power multi-channel electronics readout, improved calibration and simulation tool-chains, and data driven analysis, such as AI/ML methods. Beyond fundamental science, antineutrino detection is increasingly discussed in applied contexts, including remote monitoring, nonproliferation, and emerging maritime nuclear concepts.

Critically, ocean-bottom detection is no longer only a concept study. Recent work, including student research at University of Hawai'i and multiple theses at Tohoku University, has advanced key elements. This research includes background assessments, \textsc{Geant4} simulations, directionality algorithms, prototype detector development, structural analyses, and dedicated measurements of liquid scintillator performance at deep-ocean temperatures. Planned near-term deployments of a prototype to multi-kilometer depth will directly address remaining questions about {\it in situ} performance under hydrostatic pressure and operational constraints.

We therefore propose a phased program to transform OBD design from concept to capability: (i) targeted R\&D and at-sea prototype validation with deployment to mature operational sites such as the ALOHA Cabled Observatory, retiring the principal technical and operations risks, followed by (ii) design maturation and partnership development leading to (iii) a full-scale relocatable deep-ocean detector and a multi-site measurement campaign. 
By exploiting existing deep-ocean science sites with cabled power and data transference, we reduce cost and risk factors. 
When fully developed, OBD's mobility will enable the acquisition of geoneutrino data at multiple ocean locations. 
The OBD can move beyond a single mantle-averaged constraint toward the first geoneutrino map of mantle radiogenic compositions, and provide the necessary data to test competing models for mantle composition and heat production. 
This program would establish a new observatory-class tool for both neutrino physics and Earth sciences, enabling mantle-focused geoneutrino measurements that are not achievable on land and open a pathway to transformative cross-disciplinary results.

\section*{Acknowledgment}
This work was supported by JSPS KAKENHI Grant No. JP24K00653 and the Mitsubishi Foundation Grant No. 202410053, and was performed under the auspices of the U.S. Department of Energy by Lawrence Livermore National Laboratory under Contract DE-AC52-07NA27344, LLNL-JRNL-2019880.
We thank Steve Dye for reviewing this article and for useful discussions. 
WM thanks Tohoku University. MD thanks Blue Eisen and the oceanography engineers of Station ALOHA. VL thanks RCNS at Tohoku University and LLNL for support during the mini-sabbatical, during which part of this work was carried out.

\bibliography{ref.bib}
\bibliographystyle{aipnum4-2}

\end{document}